# Wiedemann-Franz Law and Thermoelectric Inequalities: Effective *ZT* and Single-leg Efficiency Overestimation

Byungki Ryu,[1,*] Seunghyun Oh,[1] Wabi Demeke,[2] Jaywan Chung,[1] Jongho Park,[1] Nirma Kumari,[1] Aadil Fayaz Wani,[1] Seunghwa Ryu,[2] and SuDong Park[1]

1) Energy Conversion Research Center, Korea Electrotechnology Research Institute (KERI), Changwon 51543, Republic of Korea
2) Department of Mechanical Engineering, Korea Advanced Institute of Science and Technology (KAIST), Daejeon 34141, Republic of Korea

## Abstract

We derive a thermoelectric inequality in thermoelectric conversion between the material figure of merit (*ZT*) and the module effective *ZT* using the Constant Seebeck-coefficient Approximation combining with the Wiedemann-Franz law. In a P-N leg-pair module, the effective *ZT* lies between the individual *ZT* values of the P- and N-legs. In a single-leg module, however, the effective *ZT* is less than approximately one-third of the leg's *ZT*. This reduction results from the need for an external wire to complete the circuit, introducing additional thermal and electrical losses. Multi-dimensional numerical analysis shows that, although structural optimization can mitigate these losses, the system efficiency remains limited to below half of the ideal single-leg material efficiency. Our findings explain the single-leg efficiency overestimation and highlight the importance of optimizing the P-N leg-pair module structure. They also underscore the need for thermoelectric leg-compatibility, particularly with respect to Seebeck coefficients.

* Correspondence: byungkiryu@keri.re.kr

Thermoelectric generator modules (TGMs) have attracted attention for their applications to waste heat recovery and solid state cooling.[1] For conventional P-N leg-pair TGMs, two different types of single-legs are connected thermally in parallel and electrically in series.[2] Recent work has theoretically revealed the achievable best efficiency of 17.1% over ever-explored materials.[1] However, there is an efficiency gap between theory and experiment:[1] measured values rarely achieve efficiencies around 12% for the single-state structure.[3] This gap might originate from suboptimal material selection, parasitic electrical and thermal resistance at contacts and interfaces, and thermoelectric material instability under thermal, chemical, and mechanical stresses.[1,4–7]

In the meantime, the single-leg thermoelectric device has been considered for achieving higher thermoelectric efficiency. With the single-leg structure, both P-type and N-type materials are not required, and the issue of suboptimal performance due to mismatched properties between P-type and N-type materials can be avoided. The structural simplicity of single-leg devices provides significant advantages for material performance assessment as there can be negligible parasitic resistances.

However, the *single-leg device is an incomplete circuit* and requires an additional electrically conducting wire to connect the hot and cold-sides for real-world applications. In typical single-leg performance measurements, power generation and heat flow are evaluated for the leg alone. The electrical losses through the connecting wire are typically bypassed in four-probe measurement and the thermal flow is measured at the cold-side, neglecting the flow through the hot-side wire.[8]

In this work, we report results of a thermoelectric performance analysis for single-leg thermoelectric devices using the thermoelectric algebraic equation formalism within the Constant Seebeck-coefficient Approximation (CSA). We derive a *thermoelectric inequality* between the effective $ZT$ of a P-N leg-pair module and the individual material's leg $ZT$ values of the P- and N-legs. We find that the optimal effective module $ZT$, with optimized geometry, is constrained by the $ZT$ values of the constituent single-legs. Moreover, by combing the Wiedemann-Franz law to the CSA theory, we further show that the efficiency of a single-leg module with the external wire is substantially lower than its intrinsic efficiency due to unavoidable energy losses through the temperature-difference applied connecting wires.

**Figure 1(a)** shows the thermoelectric generator module consists of P-type and N-type thermoelectric legs arranged electrically in series and thermally in parallel.[2] The legs are placed between two thermally conducting substrates, contacting with hot- and cold-side temperatures, $T_h$ and $T_c$. The external wires are connected to the cold-side and the electrical circuit is completed by an external load resistance, and the power ($P$) is delivered to the load. We define $A$ and $L$ as the area and length of the leg. For P- and N-type legs, we denote each leg with subscripts 1 and 2. The module efficiency can be computed or measured from all heat inputs into the TGM ($Q_h^{(\text{TGM})}$).

**Figure 1(b)** shows the single-leg thermoelectric device contains only one thermoelectric leg (either P-type or N-type) and requires a metallic wire to connect the single leg to the external load, completing the electrical circuit. Compared to the P-N leg-pair TGM, *the wire from the hot side experiences temperature difference*. Here, the material efficiency of the single-leg can be computed or measured from the heat input only into the single-leg ($Q_h^{(\text{leg})}$), while power generation can be computed considering voltage generation via thermoelectric leg only.

Throughout this work, we use the figure of merit $ZT_{\text{mid}}$ under the thermoelectric algebraic framework within the Constant Seebeck-coefficient Model and Constant Seebeck-coefficient Approximation.[9,10] Electrical and thermal current flowing through the TGM can be described by three thermoelectric properties (TEPs): Seebeck coefficient $\alpha$, electrical resistivity $\rho$, and thermal conductivity $\kappa$. In a single-leg device made of homogeneous material, these TEPs are functions of temperature $T$. In the Constant Seebeck-coefficient Model (CSM),[9] where the Seebeck coefficient is constant as $\alpha_0$, the optimal thermoelectric efficiency $\eta_{\text{opt}}^{(\text{CSM})}$ of the leg is exactly determined by the load resistance ratio $\gamma := \frac{R_L}{R}$ and the figure of merit $ZT$:

$$\eta_{\text{opt}}^{(\text{CSM})} = \frac{\Delta T}{T_h} \cdot \frac{\sqrt{1 + ZT_{\text{mid}}} - 1}{\sqrt{1 + ZT_{\text{mid}}} + \frac{T_c}{T_h}}, \qquad (1)$$

where $T_{\text{mid}} := \frac{T_h + T_c}{2}$. Within the CSA, the CSM theory extends for non-constant Seebeck coefficient cases. Using average parameters for three TEPs, the optimal efficiency formula can be rewritten using the CSA figure of merit of the material $Z_{mat}$ as:

$$\alpha_0 = \frac{1}{\Delta T}\int_{T_c}^{T_h} \alpha(T)\,dT, \quad \kappa_0 = \frac{1}{\Delta T}\int_{T_c}^{T_h} \kappa(T)\,dT, \quad \rho_0 = \frac{1}{\kappa_0 \Delta T}\int_{T_c}^{T_h} \rho(T)\,\kappa(T)\,dT \qquad (2)$$

$$Z_{mat} := \frac{\alpha_0^2}{RK} = \frac{\alpha_0^2}{\rho_0 \kappa_0}, \qquad (3)$$

where $R$ and $K$ are electrical resistance ($\rho_0 \frac{L}{A}$) and thermal conductance ($\kappa_0 \frac{A}{L}$) of the leg for zero-current, respectively. For simplicity, we will omit the subscript 0 when referring to the average TEPs. Previously, it was numerically validated that the efficiency prediction using **equations (2) and (3)** is highly accurate for $Z_{mat}T_{\text{mid}} < 2$, with the root-means-square of the relative efficiency error being less than 11%.

In the P-N leg-pair TGM, the external wires can be thick, and the wire resistance can be neglected. The effective $ZT$ for the P-N leg-pair ($Z_{12}$) module can be generalized within the thermoelectric algebraic framework, as reported previously:[2,10]

$$Z_{12} := \frac{(\alpha_1 + |\alpha_2|)^2}{(R_1 + R_2)(K_1 + K_2)} = \frac{(\alpha_1 + \alpha_2)^2}{(R_1 + R_2)(K_1 + K_2)}. \qquad (4)$$

For simplicity, without loss of generality, we write $|\alpha_2|$ as $\alpha_2$. Let structure factor $a$ be the relative area ratio of leg 2 to leg 1: $a := \frac{A_2}{A_1}\frac{L_1}{L_2}$. The effective module figure of merit $Z_{12}$ is rewritten as:

$$Z_{12} := \frac{(\alpha_1 + \alpha_2)^2}{\left(\rho_1 + \frac{\rho_2}{a}\right)(\kappa_1 + \kappa_2 a)} = \frac{(\alpha_1 + \alpha_2)^2}{\rho_1\kappa_1 + \rho_2\kappa_2 + (\rho_1\kappa_2)a + (\rho_2\kappa_1)a^{-1}}. \qquad (5)$$

And the module $Z_{12}$ is maximized with the optimal structure factor $a = a_{opt}$, given by:

$$a_{opt} := \sqrt{\frac{\rho_2 \kappa_1}{\rho_1 \kappa_2}}, \tag{6}$$

resulting in the optimal $Z$ ($Z_{opt}$):

$$Z_{12} \leq Z_{opt} := \frac{(\alpha_1 + \alpha_2)^2}{\rho_1 \kappa_1 + \rho_2 \kappa_2 + 2\sqrt{\rho_1 \kappa_1 \rho_2 \kappa_2}} = \left(\frac{\alpha_1 + \alpha_2}{\sqrt{\rho_1 \kappa_1} + \sqrt{\rho_2 \kappa_2}}\right)^2. \tag{7}$$

Here, the optimal structure factor $a_{opt}$ represents the balance of electrical and thermal losses in the P-N leg-pair module. Note that this optimal value applies to the optimal efficiency, not for the maximum power condition. For maximum power, a larger area is needed to reduce electrical resistance. However, this also leads to a higher heat current, which results in a significant reduction in efficiency.

When P- and N-legs are identical except for the sign of the Seebeck coefficient, the effective module $ZT$ is equal to the single-leg material $ZT$. However, in most cases, the $ZT$ values of the legs differ between the P- and N-legs. Let $Z_1 = \frac{\alpha_1^2}{\rho_1 \kappa_1} \leq Z_2 = \frac{\alpha_2^2}{\rho_2 \kappa_2}$. Then, the optimal effective $ZT$ of the leg-pair module lies between them:

$$\sqrt{Z_{opt}} = \frac{\alpha_1 + \alpha_2}{\alpha_1/\sqrt{Z_1} + \alpha_2/\sqrt{Z_2}} = \sqrt{Z_1} \frac{\alpha_1 + \alpha_2}{\alpha_1 + \alpha_2 \sqrt{Z_1/Z_2}} \geq \sqrt{Z_1}, \tag{8}$$

$$\sqrt{Z_{opt}} = \sqrt{Z_2} \frac{\alpha_1 + \alpha_2}{\alpha_1 \sqrt{Z_2/Z_1} + \alpha_2} \leq \sqrt{Z_2}, \tag{9}$$

$$Z_1 \leq Z_{opt} \leq Z_2. \tag{10}$$

From **equation (9)**, we can identify the source of the incompatibility between the legs: the figure of merit ratio, $Z_2/Z_1$. To increase the optimal effective $Z_{opt}$, it is required to reduce the ratio $Z_2/Z_1$. From **equation (10)**, we learn that the optimal efficiency of the module is constrained between the performances of the individual leg materials and cannot exceed the performance of the single-leg. This emphasizes the importance of compatibility between the performance of the legs; otherwise, the module's overall performance may be significantly reduced.

We derive an approximate form of effective $ZT$ for the case of metallic materials or highly degenerate semiconductors, where the Wiedemann-Franz Law holds. We assume that thermal conductivity is primarily determined by the electrical conductivity $\sigma := \frac{1}{\rho}$, the Lorenz number $L_0$ for metallic limit, and temperature $T$, such that $k = L_0 \sigma T$. Then, each single-leg can have similar or identical values of $\rho \kappa = L_0 T$, and we obtain:

$$[RK]_{\text{metal}} = \frac{1}{\Delta T}\int_c^h \rho\kappa\, dT = \frac{1}{\Delta T}\int_c^h L_0 T\, dT = L_0 T_{\text{mid}}, \quad (11)$$

$$[Z_{mat}T_{\text{mid}}]_{\text{metal}} = \frac{\alpha^2}{L_0}, \quad (12)$$

$$\sqrt{[Z_{opt}]_{\text{metal}}} = \frac{\alpha_1 + \alpha_2}{2\sqrt{L_0 T_{mid}}}. \quad (13)$$

From **equation (13)**, we conclude that increasing or optimizing the Seebeck coefficient is the primary objective for metallic materials or degenerate semiconductors.

When Seebeck coefficients are similar ($\alpha_1 \approx \alpha_2$), the main cause of performance incompatibility between the P- and N-legs lies in the ratio of lattice thermal conductivity ($\kappa_{la}$) to electrical conductivity $\sigma$:

$$\frac{Z_2}{Z_1} \approx \frac{\rho_1 \kappa_1}{\rho_2 \kappa_2} = \frac{L_1 + \frac{\kappa_{la,1}}{\sigma_1 T_{\text{mid}}}}{L_2 + \frac{\kappa_{la,2}}{\sigma_2 T_{\text{mid}}}}. \quad (14)$$

where $L_{1,2} = \frac{\kappa_{el}}{\sigma T_{\text{mid}}}$ is calculated for each leg. Note that, for simplicity, the subscript "mid" is omitted from this point onward in the derivation. The Seebeck coefficient and electrical conductivity are both related to the carrier concentration. However, the carrier concentration for optimal Seebeck coefficient may differ from that for the thermal-to-electrical conductivity ratio, leading to value differences that make optimizing P-N leg-pair modules challenging.

The single-leg material must be connected to an external metallic wire; otherwise, it remains electrically disconnected. The key difference between P-N leg-pair modules and single-leg modules lies in the temperature differences across the external wires. In P-N leg-pair modules, the wires are connected from the cold-side and thick wires give negligible Joule heat loss. In contrast, the hot-side wire in the single-leg introduces additional heat flow through metallic wire outside the single-leg. To minimize this heat flow, a thick metallic wire should be avoided. However, using a thinner wire increases the Joule heating in the system due to high wire electrical resistance. Therefore, it is necessary to optimize the thickness of the hot-side wire, similar to the optimization process for conventional P-N leg-pair modules.

Let leg 1 in **equations (4-7)** be a semiconducting single-leg, while leg 2 represents the metallic wire. Both thermoelectric and metallic materials follow the Wiedemann-Franz Law. For legs 1 and 2, we have:

$$\rho_1 \kappa_1 = \frac{\kappa_{la} + \kappa_{el}}{\sigma_1} = L_1 T + \frac{\kappa_{la}}{\sigma_1}, \tag{15}$$

$$\rho_2 \kappa_2 = L_0 T, \tag{16}$$

where $\kappa_{el}$ is the electronic thermal conductivity. For good thermoelectric materials, the contributions of lattice thermal conductivity and electronic thermal conductivity are comparable.

Within the parabolic band, energy-dependent relaxation time, and rigid band approximation, the Lorenz number can be calculated for metallic limit as $L_0$ and semiconductor limit as $L_1$:[2]

$$L_0 = \frac{\pi^2}{3}\left(\frac{k}{e}\right)^2 = 2.44 \times 10^{-8} \text{ V}^2 \cdot \text{K}^{-2} \tag{17}$$

$$L_1(r) = \left(\frac{k}{e}\right)^2 \left(r + \frac{5}{2}\right) \tag{18}$$

$$L_1\left(r = -\frac{1}{2}\right) = 1.49 \times 10^{-8} \text{ V}^2 \cdot \text{K}^{-2} \approx 60\% \times L_0 \tag{19}$$

where $k$, $e$, and $r$ are Boltzmann constant, electronic charge, and scattering exponent for energy-dependent relaxation time. The dominant scattering is assumed to be electron-acoustic interaction with $r = -\frac{1}{2}$, and it gives the $L_1$ of 60% of the metallic limit. Meanwhile, the Seebeck coefficient of metals are negligible compared to semiconductors, typically below $10 \text{ μV} \cdot \text{K}^{-1}$ for Cu, Ag, and Au.[13] Therefore, for semiconductor single-leg cases, we may have:

$$\rho_1 \kappa_1 = L_1 T + \frac{\kappa_{la}}{\sigma_1} \sim 2L_1 T \sim 1.2 \times L_0 T = 1.2 \times \rho_2 \kappa_2 , \tag{20}$$

$$Z_{opt}^{(single)} = \frac{(\alpha_1 + \alpha_2)^2}{\left(\sqrt{\rho_1 \kappa_1} + \sqrt{L_0 T_{\text{mid}}}\right)^2} \sim \frac{(\alpha_1 + 0)^2}{\left[\left(1 + \frac{1}{\sqrt{1.2}}\right) \times \sqrt{\rho_1 \kappa_1}\right]^2} \approx \frac{Z_{mat}}{3.7} < \frac{Z_{mat}}{3} . \tag{21}$$

where we name the inequality in **equation (21)** as ***thermoelectric single-leg inequality***.

The thermoelectric single-leg inequality provides physical insight into the overestimation of single-leg material efficiency. For $T_c = 300 \, K$, $T_c = 800 \, K$ and $Z_{mat} T_{\text{mid}} = 1$, the ideal single-leg efficiency is 17.9% using the CSA efficiency formula. However, considering the system efficiency using **equation (21)**, it is 6.6% as $[Z_{opt} T_{\text{mid}}]_{\text{leg-wire}} \sim \frac{Z_{mat} T_{\text{mid}}}{3.7} < \frac{1}{3}$. Thus, the module efficiency for the single-leg is estimated to be less than half of the single-leg material efficiency. The result highlights the importance of the P-N leg-pair module structure with a counter-leg material.

We perform one-dimensional (1D) thermoelectric integral formalism to compute numerical efficiency of single-leg and P-N leg-pair modules, within the framework of the three thermoelectric degrees of freedom theory.[11] The thermoelectric-integral equations are constructed using a given TEP set. Then, the temperature distribution is computed from an initial temperature profile under specified electrical and thermal conditions. The temperature distribution is subsequentially updated via Picard iteration, using the thermoelectric integral equations. Once the temperature converges, the device parameters, electrical power, and the heat flux are calculated. For three-dimensional (3D) simulations, we solve the temperature and voltage solutions of thermoelectric-differential equations using finite-element method, implemented in COMSOL software.[12] The electrical and thermal boundary conditions are Dirichlet type, with the current being specified. For the single-leg module, a copper wires are connected to both the hot- and cold-sides of the single-leg, following experimental measurement setting.[8] See **SM and Figure S1** for detailed geometries and boundary conditions. For all calculations, the TEPs are piecewise-linear-interpolated and constant-extrapolated. The optimal efficiency, defined as the maximum of the thermoelectric conversion efficiencies, is determined by searching for the optimal current.

We numerically validate ***thermoelectric single-leg inequality*** in **equation (21)** by performing 1D thermoelectric efficiency calculations for a single-leg wired with Cu metal when $T_c = 323\ K$ and $T_h = 573\ K$. We use the TEPs of the Bi$_{0.5}$Sb$_{1.5}$Te$_3$–Ag0.05wt% alloy (BiSbTe),[14] for the single-leg material (**see Table S1 in SM**). For Cu, we follow the electrical conduction definition of the International Annealed Copper Standard (IACS), using properties at 20°C: for 100% IACS, the electrical resistivity is 1.7241 μΩ·cm, with a temperature coefficient of 0.00393 K$^{-1}$.[15] For thermal conductivity of Cu, we use 384 W·m$^{-1}$·K$^{-1}$. We use the thermoelectric single-leg size of $3 \times 3 \times 3\ mm^3$. The Cu wire length is considered as 3 mm for 1D, and 3-7.5 mm for 3D, while the area of hot-side Cu wire ($a_{Cu}$) is considered as:

$$10^{-6} \times a_{leg} \leq a_{Cu} \coloneqq \frac{\pi}{4}D^2 \leq 1 \times a_{leg} \tag{22}$$

where $D$ is the diameter of the hot-side Cu wire. The cold side wire diameter is set to 1 mm.

**Figure 2** shows the maximum TGM efficiencies for the BiSbTe single-leg with various Cu wire diameters. When the Cu wire is very thick, power loss is negligible, whereas a thin Cu wire acts as a resistor, causing Joule heat loss, and the power vanishes: see **Figure 2(a)**. In contrast, the thin Cu wire results in a smaller heat current, while a thick Cu wire produces a large heat current: see **Figure 2(b)**. Interestingly, the heat current at the optimal current for optimal efficiency is smaller for a thin Cu wire than for a thick one. This is due to the increased system resistance from the thin Cu wire, which leads to a corresponding decrease in the optimal current for maximum efficiency. As the optimal current decreases, the Peltier heat current contribution to the total heat input becomes smaller. However, as the Cu diameter increases, thermal conduction through the Cu significantly increases and the Peltier heat current in the leg increases. Overall, as seen in **Figure 2(c) and (d)**, the module efficiency of the single-leg with a metallic connecting wire is significantly lower than that of the ideal single-leg. For the single-leg, the material efficiency is 8.97%. However, when accounting for the effects of Joule heating and thermal conduction through the metallic wire, the module efficiency drops to 3.52%. The ratio of the effective $ZT$ of the single-leg module with the wire to the material $ZT$ of the leg without the wire varies with the Cu wire diameter due to the variation of electrical resistance and thermal conductance. The effective ZT of the module is maximized to 0.292 when the wire diameter is 0.151 mm, compared to the material ZT of 0.839. The $ZT$ and efficiency ratios of the module to the material are maximized as 35% and 43%, respectively .

As shown in **Table 1 and Figure 2**, both 1D and 3D simulations demonstrate single-leg efficiency overestimation when wires are not included in performance measurement, as predicted by **equation (16),** supporting our theoretical formulation of *the thermoelectric single-leg inequality*. This overestimation can be explained because of the neglect for the temperature difference applied in the external wires, which results in *electrical and thermal energy loss* in the three-dimensional single-leg TGM measurement. The non-negligible voltage and power drop is observed owing to the wire resistance: see **Figures 3(a), S2(a), and S2(b)**. As the temperature difference exists, a large thermal flux flows through the wire from the hot side: see **Figures 3(b) and Figure S2(c)**. Increasing diameter can cause an increase in thermal energy loss by thermal conduction. Decreasing diameter can cause an increase of electrical energy loss by Joule heating. One may optimize the wire size for optimal efficiency, but it is still below half of the ideal material efficiency. This is unavoidable because the temperature difference must be applied. One may imagine that wire has no temperature difference. However, it will cause a temperature difference across the external load. Again, the situation is the same and the Wiedemann-Franz law induces thermoelectric inequality, and the module efficiency drop. There can be other sources of energy loss in the single-leg system efficiency such as Seebeck coefficient of metallic wires, reduced effective temperature difference across the single-leg, and parasitic resistances in the measurement system. As they can be controlled with a long-leg geometry, the main lesson in this work is still valid: **see Supporting Material**.

Our results emphasize the importance of the P-N leg-pair structure in the thermoelectric conversion module. In the single-leg structure, the metallic wire acts as a counter leg, leading to significant electrical and thermal losses, which cause a drop in efficiency. Although the single-leg efficiency measurements at the cold-side provide a simpler method and suggest higher conversion efficiency, the actual system-level module efficiency is much lower. To overcome this limitation, we conclude that the counter leg must also have thermoelectric performance with an opposite Seebeck coefficient. This underscores the importance of simultaneously searching for P- and N-leg materials.

In summary, we develop the theory for the effective $ZT$ for the P-N leg-pair module within the Constant Seebeck-coefficient Approximation. This theory is applied for the single-leg case with the Wiedemann-Franz law. We find that the module efficiency of the single-leg TGM is quite lower than the leg efficiency of the single-leg material owing to electrical and thermal losses through the hot-side wire. This explains the single-leg efficiency overestimation. It also emphasizes the importance of electrical circuit completeness and the compatibility between P- and N-legs in the design of advanced thermoelectric modules.

# Table

**Table 1.** 1D and 3D simulation of thermoelectric performance for BiSbTe single-leg without and with Cu external wire. Thermoelectric performances and efficiencies are calculated at the optimal current condition ($I_{opt}$) for maximum efficiency: also see **Table S2** in **SM**.

| Model | Connecting Cu wire(s) (length L, diameter D) | Measure | $Z_0 T_{mid}$ | $I_{opt}$ (A) | $P$ (mW) | $Q_h$ (mW) | $\eta_{opt}$ |
|---|---|---|---|---|---|---|---|
| 1D-1 | No Cu wire | Single-leg | 0.839 | 3.078 | 90.0 | 1102 | 8.17% |
| 1D-2 | Hot wire (3 mm, 0.151 mm) | TGM | 0.292 | 2.045 | 55.4 | 1571 | 3.52% |
| 3D-1 | Hot wire (3 mm, 0.151 mm), Cold wire (3 mm, 1 mm) | Single-leg | - | 3.034 | 87.8 | 1086 | 8.09% |
| 3D-1 | Hot wire (3 mm, 0.151 mm), Cold wire (3 mm, 1 mm) | TGM with wire | - | 2.040 | 54.7 | 1589 | 3.44% |
| 3D-2 | Hot wire (7.5 mm, 0.239 mm), Cold wire (7.5 mm, 1 mm) | TGM with wire | - | 1.996 | 53.5 | 1565 | 3.42% |

# Figure Captions

**Figure 1.** Schematic view of (a) the P-N leg-pair TGM and (b) the single-leg TGM structure.

**Figure 2.** 1D optimal thermoelectric performance of the BiSbTe single-leg with various Cu wire diameters. (a) Power generation, (b) inverse heat input at $T_h$, (c) efficiency, and (d) *ZT* and maximum efficiency ratio between with and without wires.

**Figure 3.** Thermoelectric performance of the BiSbTe single-leg TGM of Model 3D-2 is visualized for the electrical current of 1.996 A and the Cu wire lengths of 7.5 mm (see Table 1 for Model). (a) Current-density vector arrows and voltage iso-surfaces. (b) Thermal-flux vector arrows and temperature iso-surfaces.

# Acknowledgements

This work was supported by the Primary Research Program of KERI through the National Research Council of Science & Technology (NST) funded by the Ministry of Science and ICT (MSIT) (grant no. 24A01019), and by the Korea Institute of Energy Technology Evaluation and Planning (KETEP) grant funded by the Ministry of Trade, Industry and Energy (MOTIE) (grant no. 2021202080023D). It was also partially supported by the National Research Foundation of Korea (NRF) grant funded by the Korea Government (MSIT) (grant no. 2022M3C1C8093916).

# Conflict of Interest Statement

The authors have no conflicts to disclose.

# Author Contributions

All authors contributed to writing, reviewing, and editing the manuscript. BR conceptualized and investigated the study. BR and JC developed the 1D software. SDP led the project. SO investigated the study for 3D simulations, under the guidance of BR and WD. BR, JC and SDP acquired funding. JP and NK discussed the module measurement. BR, SO, WD, JC, NK, AFW, SR, and SDP contributed to discussion on efficiency theory.

# Data Availability Statement

The data that support the findings of this study are available from the corresponding author upon reasonable request.

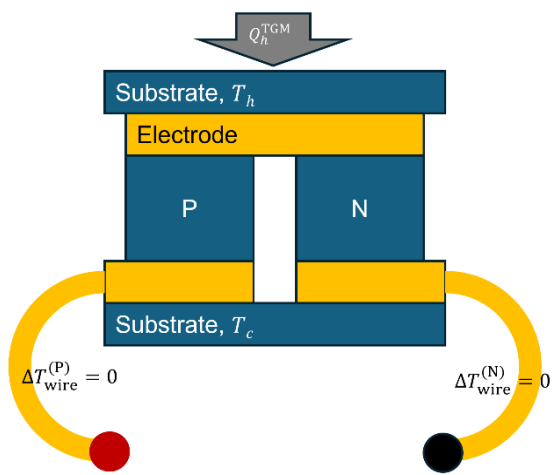 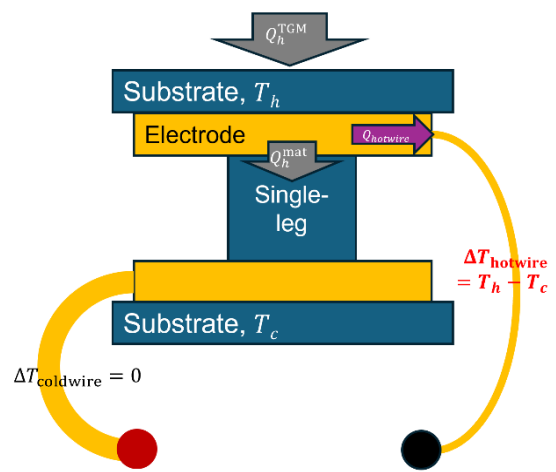

**Figure 1**

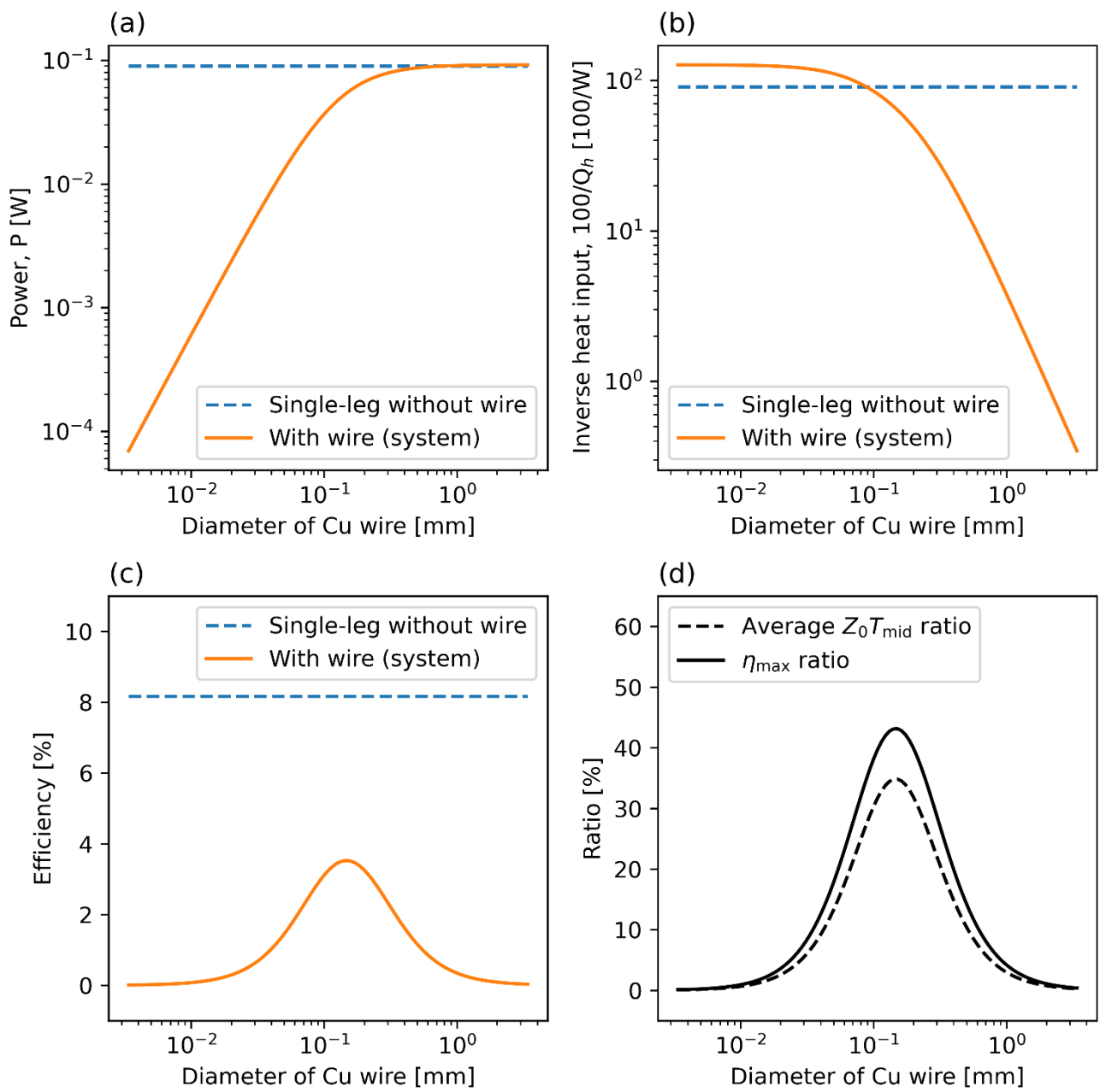

**Figure 2**

**(a)**

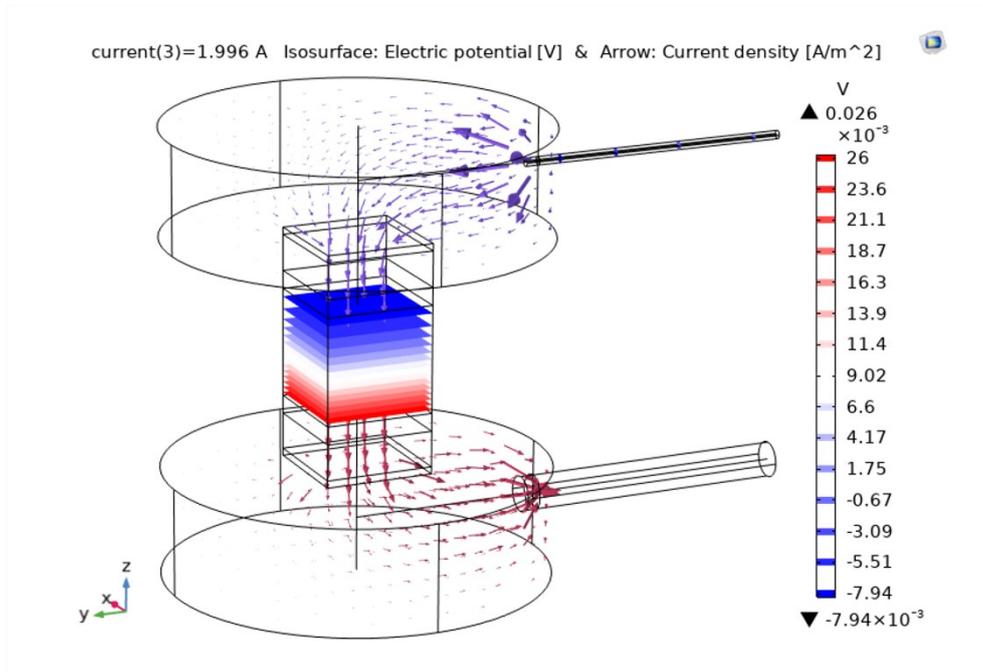

**(b)**

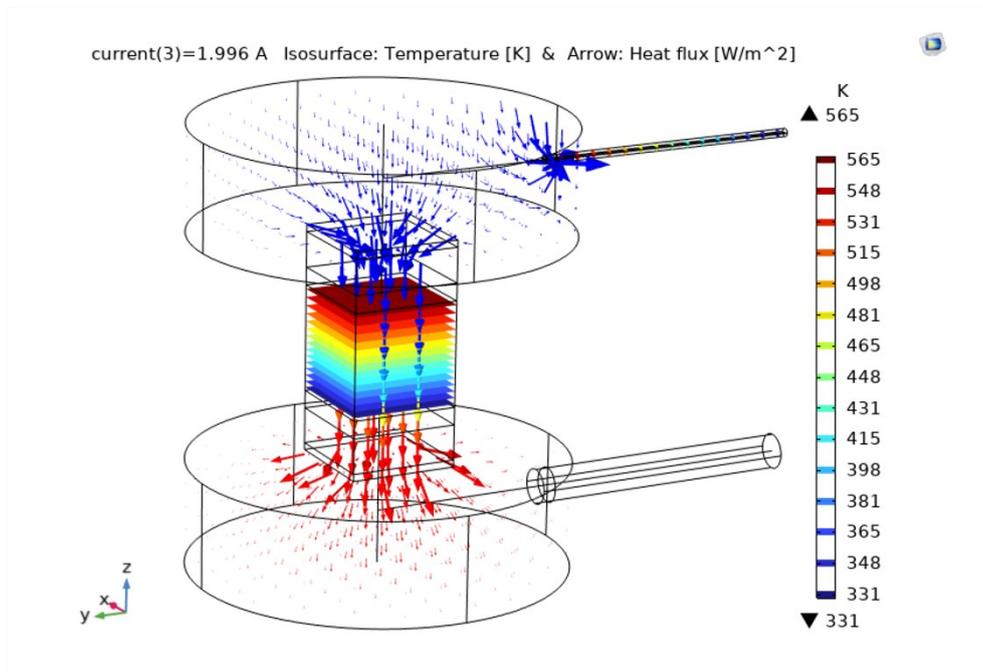

**Figure 3**

# Supporting Material for "Wiedemann-Franz Law and Thermoelectric Inequalities: Effective *ZT* and Single-leg Efficiency Overestimation"

Byungki Ryu,[1,*] Seunghyun Oh,[1] Wabi Demeke,[2] Jaywan Chung,[1] Jongho Park,[1] Nirma Kumari,[1] Aadil Fayaz Wani,[1] Seunghwa Ryu,[2] and SuDong Park[1]

1) Energy Conversion Research Center, Korea Electrotechnology Research Institute (KERI), Changwon 51543, Republic of Korea
2) Department of Mechanical Engineering, Korea Advanced Institute of Science and Technology (KAIST), Daejeon 34141, Republic of Korea

## A. Detailed Geometry the Three Dimensional Single-Leg Structure

See Figure S1(a) for the geometry and related parameters. See Figure S1(b) for planes for the voltage generated and input heat measurement. Note that the cold temperature is also applied otherwise the heat will damage the outside system.

## B. Efficiency Loss in the Three Dimensional Single-Leg Structure

There can be additional origin of energy and efficiency loss in the single-leg structure in addition to the analysis on the main manuscript, but they are found to be negligible.

**Effective temperature difference change.** Another origin of efficiency drop is the effective temperature difference change along the single-leg due to the existence of substrate. However, as the thermal conductivity of the Cu substrate is very large, the temperature difference loss is very small compared to the one-dimensional result: about 1% of $\Delta T = 250\ K$. Thus, the efficiency drop by substrate is negligible.

**Wire Seebeck coefficient.** Every metal element has non-zero Seebeck coefficient. However, our theoretical and computational analysis is for the zero Seebeck-coefficient metallic wire. Compared to the system efficiency, where the single-leg is connected to the external wire, the material $ZT$ and single-leg efficiency measurements highly overestimate the thermoelectric performance. This implies that if we include the wire's Seebeck coefficient, the single-leg efficiency could be more erroneous. The Seebeck coefficient of BiSbTe we used is about 200 $\mu V \cdot K^{-1}$, while the experimental Cu Seebeck coefficient is around 1.5 $\mu V \cdot K^{-1}$ at room temperature, approximately 1% of that of the thermoelectric material. However, Fe and $d^8$ transition metal show a non-negligible Seebeck coefficient of approximately 10 $\mu V \cdot K^{-1}$.[13] Given that the $ZT$ depends on the square of the Seebeck coefficient, the error in $ZT$ can reach about 10%. If the metallic Seebeck coefficient is negative while the single-leg is P-type, the error becomes even more significant, potentially leading to a 10% higher $ZT$ value in the single-leg measurement, thereby further increasing the efficiency overestimation.

**Parasitic interfacial and contact resistances.** Another issue is the presence of parasitic resistance, which significantly degrades thermoelectric efficiency. In the single-leg system, due to the simplicity of leg fabrication, the electrical and thermal contact or interfacial resistances are lower than those in P-N leg-pair modules. Therefore, there will be a noticeable performance gap between single-leg devices and standard thermoelectric modules.

## C. Thermoelectric properties

We use the thermoelectric properties of the $Bi_{0.5}Sb_{1.5}Te_3$–$Ag0.05wt\%$ alloy (BiSbTe) [Ref] for the single-leg material.

    **[Ref]** J. K. Lee et al. Electron. Mater. Lett. 6, 201 (2010). https://doi.org/10.3365/eml.2010.12.201.

    In KERI TE team, the material DB code and label are as below. The DB will be open in future.

        [sampleid] 43

        [dbinfo] tematdb_v1.1.0_completeTEPset_convertedOn_20230817_204017__range_1_to_424

**Table S1.** Thermoelectric properties used for the single-leg module simulation.

| tepname | Temperature | tepvalue | unit | DB |
|---|---|---|---|---|
| alpha | 322.597 | 0.0002006830 | [V/K] | tematdb |
| alpha | 372.468 | 0.0002143340 | [V/K] | tematdb |
| alpha | 423.117 | 0.0002184300 | [V/K] | tematdb |
| alpha | 472.987 | 0.0002116040 | [V/K] | tematdb |
| alpha | 523.636 | 0.0001924910 | [V/K] | tematdb |
| alpha | 572.727 | 0.0001651880 | [V/K] | tematdb |
| rho | 322.500 | 0.0000121778 | [Ohm-m] | tematdb |
| rho | 372.500 | 0.0000154366 | [Ohm-m] | tematdb |
| rho | 423.333 | 0.0000195714 | [Ohm-m] | tematdb |
| rho | 473.333 | 0.0000233192 | [Ohm-m] | tematdb |
| rho | 524.167 | 0.0000260952 | [Ohm-m] | tematdb |
| rho | 575.000 | 0.0000267317 | [Ohm-m] | tematdb |
| kappa | 322.923 | 0.9960280000 | [W/m/K] | tematdb |
| kappa | 373.168 | 0.9397000000 | [W/m/K] | tematdb |
| kappa | 423.431 | 0.9454410000 | [W/m/K] | tematdb |
| kappa | 472.930 | 1.0201700000 | [W/m/K] | tematdb |
| kappa | 522.451 | 1.1707500000 | [W/m/K] | tematdb |
| kappa | 572.762 | 1.3351100000 | [W/m/K] | tematdb |
| ZT | 322.826 | 1.0829600000 | [1] | tematdb |
| ZT | 371.739 | 1.2040400000 | [1] | tematdb |
| ZT | 422.283 | 1.0964100000 | [1] | tematdb |
| ZT | 472.826 | 0.9080720000 | [1] | tematdb |
| ZT | 523.370 | 0.6457400000 | [1] | tematdb |
| ZT | 573.098 | 0.4573990000 | [1] | tematdb |

**Table S2.** It contains detailed information regarding the single-leg TGM model in **Table 1**. The leg power satisfies: $P^{(\text{leg})} = Q_h^{(\text{leg})} - Q_c^{(\text{leg})}$. Meantime, the TGM power satisfies: $P^{(\text{TGM})} = \left(Q_h^{(\text{TGM})} - Q_c^{(\text{TGM})}\right) - Q_{loss}^{(\text{wire})}$, where $Q_{loss}^{(\text{wire})} = Q_{h,loss}^{(\text{wire})} + Q_{c,loss}^{(\text{wire})}$. Note that in 1D simulation, the Cu wire acts as a counter-leg in the TGM so that the heat loss through the wire is included in the cold-side heat of TGM already. Also note that the size of the energy balance error in 3D simulation ($\delta_{\text{err}}$), defined as $\delta_{\text{err}} = \left(\frac{\Delta E}{Q_h^{(\text{TGM})}}\right) = \left[\frac{\left(Q_h^{(\text{TGM})} - Q_c^{(\text{TGM})}\right) - \left(Q_{loss}^{(\text{wire})} + P^{(\text{TGM})}\right)}{Q_h^{(\text{TGM})}}\right]$, is less than 0.1%.

| Model | $I$ (A) | $P^{(\text{leg})}$ | $Q_h^{(\text{leg})}$ | $Q_c^{(\text{leg})}$ | $P^{(\text{TGM})}$ | $Q_h^{(\text{TGM})}$ | $Q_c^{(\text{TGM})}$ | $Q_{h,loss}^{(\text{wire})}$ | $Q_{c,loss}^{(\text{wire})}$ | $\eta^{(\text{leg})}(I)$ | $\eta^{(\text{TGM})}(I)$ | $\delta_{\text{err}}$ |
|---|---|---|---|---|---|---|---|---|---|---|---|---|
| | | (mW) | | | | | | | | | | |
| 1D | 3.08 | 90.0 | 1102 | 1012 | - | - | - | - | - | 8.17% | - | - |
| | 2.05 | 74.9 | 1005 | 930 | 55.4 | 1572 | 1516 | - | - | 7.45% | 3.52% | - |
| 3D-1 | 3.034 | 87.8 | 1086 | 998 | 46.1 | 1670 | 997 | 625 | 3 | 8.09% | 2.76% | -0.07% |
| | 2.040 | 73.5 | 993 | 919 | 54.7 | 1589 | 918 | 615 | 2 | 7.34% | 3.44% | -0.06% |
| 3D-2 | 1.996 | 72.5 | 988 | 916 | 53.5 | 1565 | 915 | 595 | 1 | 7.34% | 3.42% | +0.04% |

## (a) 3D single-leg system structure

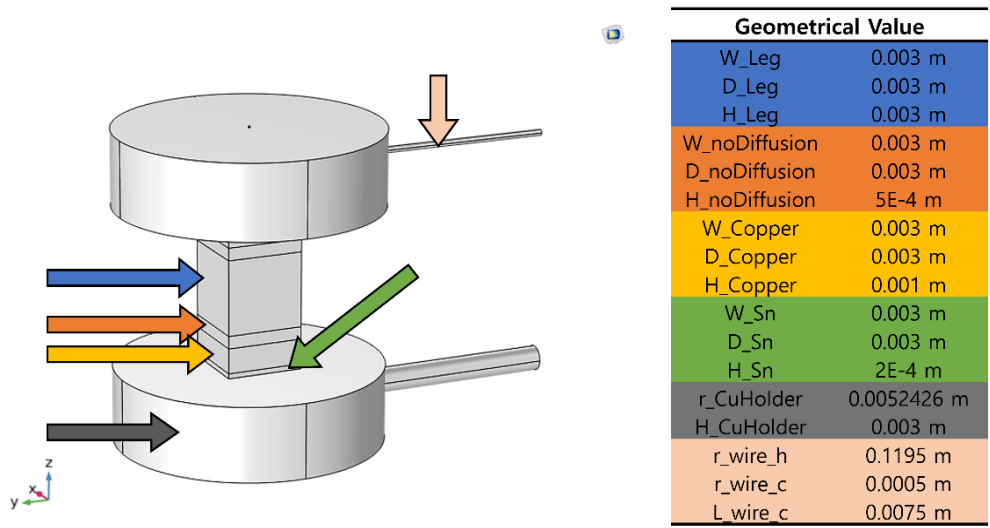

| Geometrical | Value |
|---|---|
| W_Leg | 0.003 m |
| D_Leg | 0.003 m |
| H_Leg | 0.003 m |
| W_noDiffusion | 0.003 m |
| D_noDiffusion | 0.003 m |
| H_noDiffusion | 5E-4 m |
| W_Copper | 0.003 m |
| D_Copper | 0.003 m |
| H_Copper | 0.001 m |
| W_Sn | 0.003 m |
| D_Sn | 0.003 m |
| H_Sn | 2E-4 m |
| r_CuHolder | 0.0052426 m |
| H_CuHolder | 0.003 m |
| r_wire_h | 0.1195 m |
| r_wire_c | 0.0005 m |
| L_wire_c | 0.0075 m |

## (b) 3D single-leg measurement

**Voltage measurement for electrical performance**

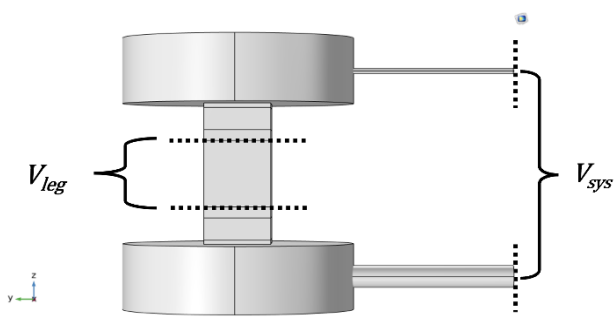

**Heat input ($Q_h$) measurement for thermal performance**

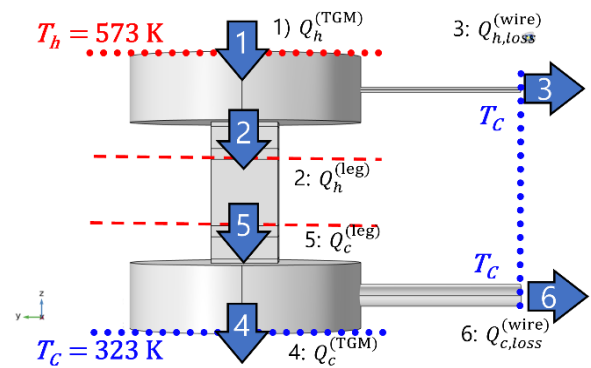

**Figure S1.** 3D model for the Thermoelectric Generator Module of the single-leg with hot and cold wires. (a) 3D geometry and parameters. (b) Visualization of the electrical and thermal boundary conditions, and the voltage and heat input measure planes for electrical and thermal performance analysis for the single-leg module with wires.

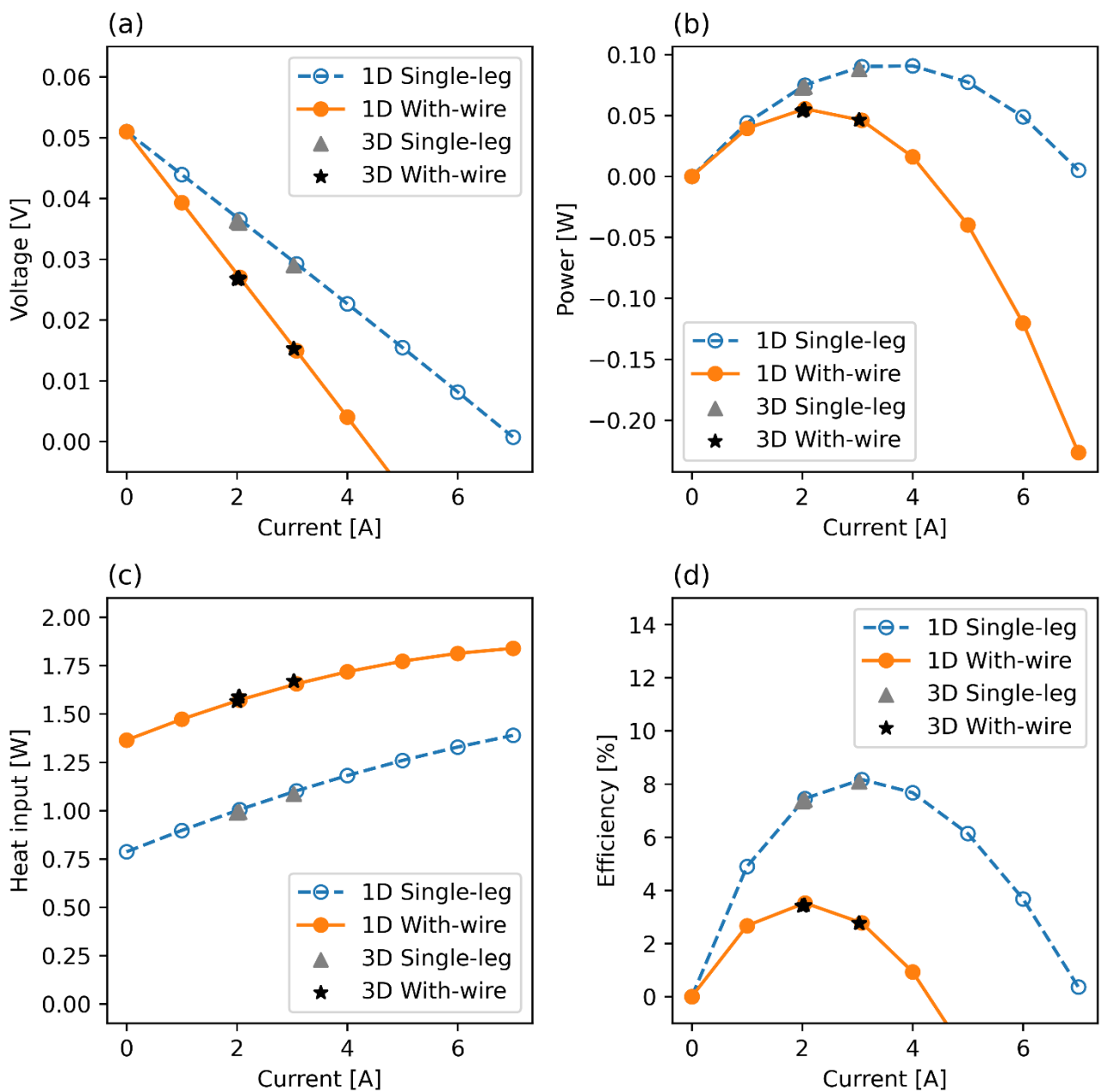

**Figure S2.** 1D and 3D thermoelectric performances of the BiSbTe single-leg based TGMs with the optimal hot-side Cu wire geometry. (a) Voltage, (b) net power generation, (c) hot side heat input, and (d) efficiency are computed with various currents. The 3D results are calculated from the FEM COMSOL simulation, while 1D results from 1D integral formalism simulation. The 3D simulation results are close to the 1D simulation results, indicating that the theory of thermoelectric inequality can be applied to explain single-leg efficiency overestimation.